 \documentclass[final,5p,times,twocolumn]{elsarticle}

\usepackage{amssymb}
\usepackage{amsmath}
\usepackage{lipsum}
\usepackage{graphicx}
\usepackage{comment}

\newcommand{\bea}{\begin{eqnarray}}
\newcommand{\eea}{\end{eqnarray}}
\newcommand{\be}{\begin{equation}}
\newcommand{\ee}{\end{equation}}

\newcommand{\re}[1]{(\ref{#1})}

\DeclareMathOperator{\Tr}{Tr}

\journal{Physics Letters B}

\begin{document}

\begin{frontmatter}



\title{Large solitons flattened by small quantum corrections}


\author[first]{Eduard Kim}
\affiliation[first]{organization={Moscow Institute of Physics and Technology},
            addressline={Institutsky lane 9}, 
            city={Dolgoprudny},
            postcode={141700}, 
            state={Moscow region},
            country={Russia}}
\author[second]{Emin Nugaev}
\affiliation[second]{organization={Institute for Nuclear Research of RAS},
            addressline={prospekt 60-letiya Oktyabrya 7a}, 
            city={Moscow},
            postcode={117312}, 
            country={Russia}}

\author[third,fourth,fifth]{Yakov Shnir}
\affiliation[third]{organization={BLTP, JINR},
            addressline={6 Joliot-Curie St}, 
            city={Dubna},
            postcode={141980}, 
            state={Moscow region},
            country={Russia}}
\affiliation[fourth]{organization={Hanse-Wissenschaftskolleg},
            addressline={Lehmkuhlenbusch 4}, 
            city={Delmenhorst},
            postcode={27733}, 
            state={Lower Saxony},
            country={Germany}}
\affiliation[fifth]{organization={Institute of Physics, Carl von Ossietzky University Oldenburg},
            addressline={Carl-von-Ossietzky-Straße 11}, 
            city={Oldenburg},
            postcode={26129}, 
            state={Lower Saxony},
            country={Germany}}

\begin{abstract}
We propose a general form of the UV-completed Friedberg-Lee-Sirlin (FLS) model. As can be seen from the mechanical interpretation, UV-completion allows thin-wall approximation for non-topological solitons. The 1-loop renormalized effective potential for the UV-completed FLS model is constructed under the assumption of mass hierarchy. By special choice of parameters, we studied how the Coleman-Weinberg mechanism induces a new mass scale $\omega_{-}$ in the classical FLS model. It can be expounded in terms of a stable condensate. As a consequence, the asymptotic of a non-topological soliton's energy at large charge is changed from $\sim Q^{3/4}$ to the linear law.
\end{abstract}




\end{frontmatter}




\section{Introduction}
\label{introduction}

Many nonlinear physical systems support solitons, spatially localized field configurations with important  ramifications for cosmology, high energy physics, non-linear optics, condensed matter and nuclear physics, and other disciplines; e.g. \cite{bishop1980solitons,dauxois2006physics,manton2004topological,shnir2018topological,vilenkin1994cosmic,manton2022skyrmions}. There are two distinct classes of solitons in field theory: (i) Topological solitons, like vortices \cite{Nielsen:1973cs}, Skyrmions \cite{Skyrme:1962vh}, instantons \cite{Belavin:1975fg} and monopoles \cite{tHooft:1974kcl,Polyakov:1974ek}, which are absolutely stable due to the topological properties of the model; (ii) Non-topological solitons such as Q-balls \cite{Rosen:1968zwl, COLEMAN1985263}, which appear due to the balance between the effects of nonlinearity and dispersion.

Recently, there has been a lot of attention to Q-balls, and it was conjectured that such configurations may be formed in a primordial phase transition, contributing to a diverse scenario of the evolution of the early Universe \cite{PhysRevD.40.3241,Kusenko:1997hj}. In particular, Q-balls were suggested to play a catalysing role in baryogenesis \cite{Affleck:1984fy, Enqvist:1997si}; it is also possible for them to occur as a form of the cold dark matter \cite{Kusenko:1997si,PhysRevLett.80.3185,Krylov:2013qe}. The Q-balls typically arise in supersymmetric generalizations of the Standard Model with the Noether charge $Q$ being identified with the baryon or lepton number \cite{Enqvist:1997si,KUSENKO1997108,PhysRevD.77.043504,Hartmann:2012gw,Campanelli:2009su}.

Both topological and non-topological solitons emerge as solutions of classical field equations, in most cases, they are very heavy, compared to the masses of perturbative excitations in the weak coupling limit. Therefore, consideration of the solitons in quantum theory typically involves semiclassical consideration \cite{Dashen:1975hd,Faddeev:1977rm}. This approach works for many models, well-known examples are the evaluation of quantum corrections to the classical mass of the kink \cite{Dashen:1974cj}, the string tension of the vortex \cite{Baacke:2008sq} and the mass of the non-Abelian monopole \cite{Kiselev:1988gf,Zarembo:1995am}. It has been pointed out that quantum corrections may strongly affect the vacuum via the Coleman-Weinberg mechanism \cite{Coleman:1973jx}, generated in such a way that effective potential may stabilize vortices \cite{Eto:2022hyt} and decrease the mass of the monopole \cite{Kiselev:1990fh}. Quantum corrections may also affect the stability of non-topological solitons \cite{Farhi_1998, Xie:2023psz}.

In general, stationary non-topological solitons are stabilized by conserved Noether charge, which can also be interpreted as the particle number. The Q-balls can be viewed as condensates of a large number of the field quanta. These field configurations correspond to the constrained extremum of the energy functional for a fixed value of the charge.

There are also  certain restrictions on the potential $V(|\phi|)$ of the model, which supports Q-balls (with standard ansatz $\phi(t,\Vec{x})=e^{-i\omega t}f(\Vec{x})$, so that $|\phi|=f$) \cite{COLEMAN1985263}. First, localized configurations must have an appropriate asymptotic at the spatial infinity, the potential must be attractive.  On the other hand, it is necessary that the potential satisfy the relation 
\be
V^{\prime \prime}(0) > \underset{\phi}{\rm min} \, [2 V(|\phi|)/|\phi|^2] \, ,
\label{Umin}
\ee
where the minimum is taken over all values of $\phi$. In addition, the hint of Q-ball's existence can be seen from the stability analysis of the condensate in the theory. The potential should satisfy the relation
\be
    V^{\prime \prime}(f) - \frac{V^{\prime}(f)}{f}<0 \, ,
    \label{1-field instability}
\ee
for some field values to maintain non-stable condensate, which is a one of the necessary conditions for Q-ball formation.

Clearly, a simplest choice of the  potential would be the renormalizable non-linear quartic self-interaction, which supplements the mass term, i.e. 
\be
V(|\phi|)= m^2 |\phi|^2 - \frac{\lambda_{\phi}}{4} |\phi|^4 \, .
\label{Extra_pot}
\ee
This choice is formally compatible with \re{Umin}, yet it does not support any stable classical solution \cite{1970JMP....11.1336A}. Additionally, the energy functional in the theory with Lagrangian \re{Extra_pot} is negatively defined, so the model demands UV completion at some scale. However, classical Q-balls may exist in a model with a single complex scalar field and a suitable nonrenormalizable self-interaction potential \cite{COLEMAN1985263,PhysRevD.39.1665,PhysRevD.66.085003}. 

The healthy model, which supports non-topological solitons in (3+1) dimensions, is the well-known two-component Friedberg-Lee-Sirlin model \cite{PhysRevD.13.2739}. 
It involves the attraction-inducing interaction between bosons of the complex field $\phi$ through an additional real scalar field $\chi$. The real field has a finite vacuum expectation value generated via a symmetry-breaking potential. The Lagrangian of the FLS model is 

\begin{equation}\label{eq.1}
            \mathcal{L} = \partial_{\mu}\phi^{\ast}\partial^{\mu}\phi + \frac{1}{2}\partial_{\mu}\chi\partial^{\mu}\chi - h^{2}|\phi|^{2}\chi^{2} - \frac{ \varkappa^{2} }{2}\left(\chi^{2} - v^{2} \right)^{2}\, ,
    \end{equation}
so, the complex field  acquires mass  due to its coupling to the real component, the bare mass $m_\phi$ is zero. 

The properties of classical soliton solutions (with standard ansatz $\phi(t,\Vec{x})=e^{-i\omega t}f(\Vec{x})$ and $\chi(\Vec{x})$) of the model \re{eq.1} and the corresponding criteria of stability were discussed in the original paper \cite{PhysRevD.13.2739}. It was pointed out that the classical FLS model can be relevant in various phenomenological applications in condensed matter and nuclear physics \cite{LEE1992251}. The coupling of this model to gravity allows the existence of boson stars and hairy black holes, which have been intensively studied in recent years \cite{Friedberg:1986tp,Liebling:2012fv,Jetzer:1991jr}. It was also used to study the production of non-topological solitons at cosmological phase transitions
\cite{Krylov:2013qe}. 

In the framework of quantum field theory, the model \re{eq.1} requires a few remarks. Firstly, it has to be considered at some renormalization scale. In the weak coupling limit, this scale is set by the masses of excitations of the fields; the corresponding quantum corrections are small. On the other hand, Q-balls can be very large, up to the values of Noether charge $Q\gtrsim 10^{14}$ \cite{Enqvist:1997si, Kusenko:1997si}. The mass of the soliton sets a different mass scale, and corresponding quantum effects may drastically change the properties of the classical solutions. Secondly, there can be some hierarchy of mass scales in the two-component model \re{eq.1}, which may allow us to generate the Coleman-Weinberg effective potential \cite{Coleman:1973jx} at one loop. Thirdly, the quantum FLS model has to be UV-completed; it should be extended by including the quantum corrections induced counterterms in the potential of the complex field.

The goal of this paper is to construct the Coleman-Weinberg potential for the FLS model under the assumption that $m_\chi \gg m_\phi$. We will start our consideration of the UV-completed FLS model in Sec.\ref{UV}. Consistent procedure of integrating-out quantum fields requires setting up an appropriate parameter hierarchy in the FLS model. We address this issue in Sec.\ref{Sec.CW}. In Sec.\ref{eff.poten.}, our goal is to derive a simplified low-energy classical theory of a single complex scalar field $\phi$ that allows Q-ball solutions. An explicit discussion of the results is given in the Outlook.

\section{UV-completed FLS model} \label{UV}
Theory (\ref{eq.1}) provides a classical example of non-topological solitons \cite{PhysRevD.13.2739}. However, UV-divergences of quantum theory require completion of
initial model (\ref{eq.1}). Assuming discrete symmetry 
$\chi\to-\chi$ general renormalizable Lagrangian has the form

\begin{equation}\label{UV FLS}
    \begin{split}
    &\mathcal{L} = \partial_{\mu}\phi^{\ast}\partial^{\mu}\phi + \frac{1}{2}\partial_{\mu}\chi\partial^{\mu}\chi - V(|\phi|, \chi), \\
    & V(|\phi|, \chi) = m_{\phi}^{2}|\phi|^{2} + \frac{\lambda}{2}|\phi|^{4} + h^{2}|\phi|^{2}\chi^{2} + \frac{\varkappa^2}{2}(\chi^{2}-v^{2})^{2}.
\end{split}
\end{equation}
The FLS model \re{eq.1} can be reproduced in the limit $m_{\phi}\to 0, \lambda \to 0$. Moreover, from a classical perspective, the model (\ref{Extra_pot}) can be regarded as an effective theory \cite{kim2023effectively} for the small-amplitude field $\phi$ of the FLS model \re{eq.1}.

The model \re{UV FLS} is itself of great interest \cite{Krylov:2013qe,Rajaraman:1975qr, Montonen:1976yk,Jiang:2023qbm, jiang2024gauged}. Firstly, we study the stability of the condensate of the model \re{UV FLS} as well as test Coleman's Q-ball conditions by using the method of the effective potential under the assumption of a heavy real field $\chi$. For two-field theories, condensate stability condition \re{1-field instability} should be modified as
\be
V_{\chi \chi}^{\prime \prime}(f,\chi)\left[V_{ff}^{\prime \prime}(f,\chi)- \frac{1}{f}V_{f}^{\prime}(f,\chi)\right] + \left[V_{f\chi}^{\prime \prime}(f,\chi)\right]^{2} <0,
\label{2-field instability}
\ee
which shows that the model \re{UV FLS} carries an unstable condensate that may form non-topological solitons. In addition, we can construct a single-field effective potential for the model \re{UV FLS} and study the Q-ball solutions. It is convenient to introduce the potential
\be
U(|\phi|) = \omega^{2}|\phi|^{2} - V_{eff}(|\phi|), 
\label{mech_pot}
\ee
where following \cite{kim2023effectively}
\be
V_{eff}(|\phi|)= V(|\phi|,\chi)\Big{\vert}_{\frac{\partial V(|\phi|,\chi)}{\partial \chi}=0}.
\label{clas_eff_pot}
\ee
Potential \re{mech_pot} can be interpreted as a mechanical potential for the spatial evolution of field $\phi$ that starts from some initial value $|\phi|_{in}$ and slides to $|\phi|=0$. 

\begin{figure}[!tbp]
  \centering
  
    \includegraphics[width=0.45\textwidth]{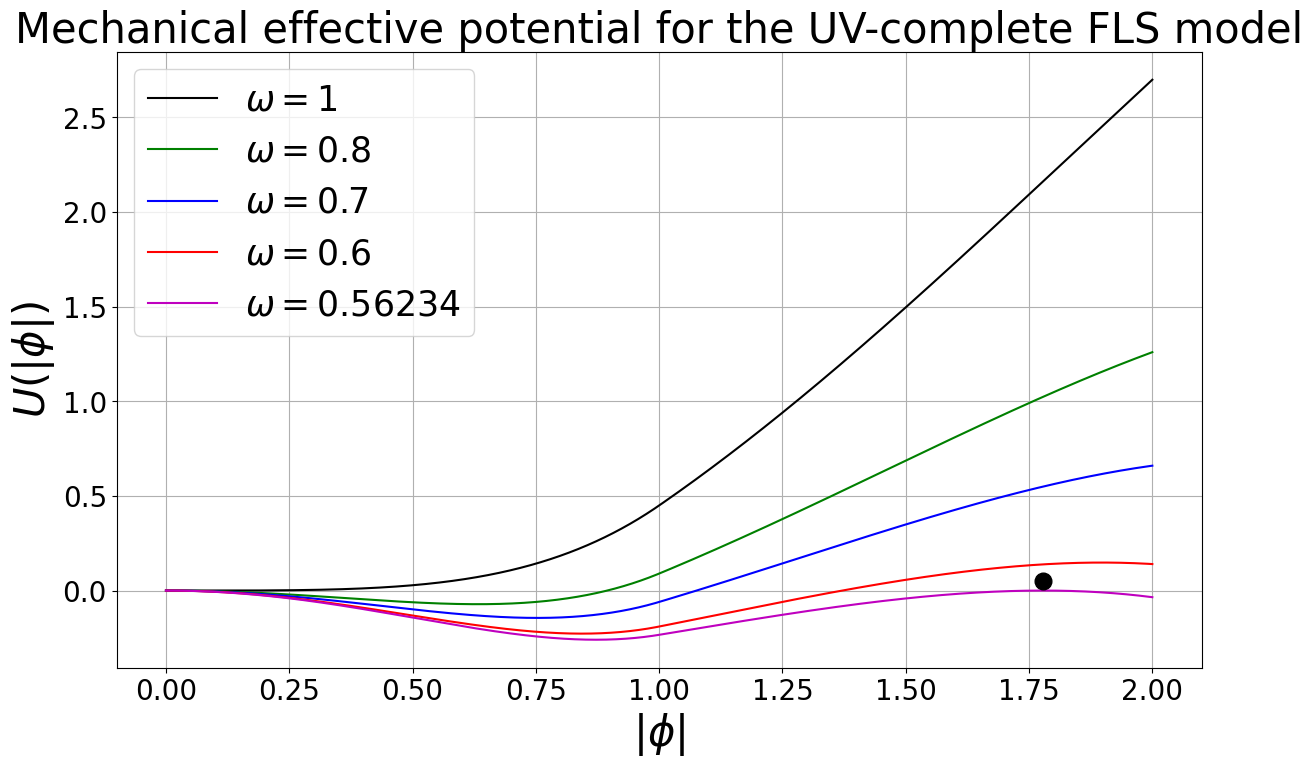}
    \caption{The mechanical potential $U(|\phi|)$ of the model \re{UV FLS} plotted for different values of parameters $\omega$. The couplings of the model are set to $h = 1$, $\varkappa = 1$ and $\lambda = 0.1$, while the nondimensionalization of the Lagrangian \re{UV FLS} is done by the substraction of the $v^{4}$ multiplier. Without loss of generality, we set $v=1$. The black dot represents a condensate.}
    \label{FLS_mech_eff}
  
\end{figure}

For a general UV-completed FLS model \re{UV FLS}, one can  find non-topological solitons numerically. We can provide an analytical consideration of the model. As mentioned above, the model \re{UV FLS} carries an unstable condensate, and now we should derive and study the effective potential of the theory. In order to simplify the calculation, we neglect the bare mass term $m_{\phi}^{2}|\phi|^{2}$ since it can only alter the mass of the linear perturbations around the vacuum and change the value of the upper limit of parameter $\omega$. By applying the gradient approximation \re{clas_eff_pot} to the equation of motion for heavy field $\chi$, we gain the integrating-out trajectories 
\be
    \begin{cases}
            & \chi = \sqrt{v^{2} - \frac{h^2}{\varkappa^{2}}|\phi|^{2}}, \text{ if } |\phi| < \frac{\varkappa v}{h}, \\
            & \chi = 0, \text{ if } |\phi| > \frac{\varkappa v}{h}.
    \end{cases}
    \label{int-out chi}
\ee

Therefore, the UV-completed FLS model results in an appropriate form of the potential \re{mech_pot} that allows for Q-ball solutions. As can be seen from Eq.\re{mech_pot}, the quartic self-interaction $\lambda$-term induces a lower bound of parameter $\omega$ (see Fig.\ref{FLS_mech_eff})
\be
\omega_{-} = v(\varkappa^{2}\lambda)^{\frac{1}{4}}, \text{ while } \lambda \leq \frac{h^{4}}{4\varkappa^{2}}.
\label{w_min_1}
\ee
One can see that while $\omega \rightarrow \omega_{-}$, the Q-ball field profile starts to resemble a thin-wall regime solution. At $\omega=\omega_{-}$, the only possible initial value of $|\phi|$ is
\be
|\phi|_{in} = C = \frac{\omega_{-}}{\sqrt{\lambda}}, 
\ee
which also provides a local extremum, so that $U(C)=0$ and $U^{'}(C) = 0$. Thus, a $\phi = C e^{-i \omega_{-} t}$ is a homogeneous solution that represents a condensate. In other words, as parameter $\omega$ approaches its lower bound $\omega_{-}$, the Q-ball's energy and $U(1)$ charge tend to infinity. These features of the UV-completed FLS model make it a healthy analogue of the non-renormalizable $|\phi|^{6}$ Q-ball model \cite{COLEMAN1985263}.

\section{Coleman-Weinberg potential}\label{Sec.CW}

In this section, we derive the 1-loop Coleman-Weinberg potential of the UV-completed FLS model \re{UV FLS} in the leading order. Following the method of the work \cite{Manohar:2020nzp}, we start by shifting fields by their classical values

\begin{equation}
    \begin{split}
        & \phi^{(\ast)} = \Hat{\phi}^{(\ast)} + \phi^{(\ast)}_{q}, \\
        & \chi = \Hat{\chi} + \chi_{q}.
    \end{split}
\end{equation}

Now, let us introduce the mass matrix $W$ \cite{Coleman:1973jx} as $W_{ij} = \frac{\partial^{2}V(\varphi)}{\partial \varphi_{q,i} \partial \varphi_{q,j}}$
for a theory with a set of $\{\varphi_{i}\}$ fields. For the FLS model \re{UV FLS}, the mass matrix is of the form 

\begin{equation}
    W=\begin{pmatrix}
m_{\phi}^{2}+\lambda |\Hat{\phi}|^{2}+h^{2}\Hat{\chi}^{2} & 2h^2\Hat{\phi}\Hat{\chi} \\
2h^2\Hat{\phi}^{\ast}\Hat{\chi} & 2h^{2}|\Hat{\phi}|^{2}+\varkappa^{2}(6\Hat{\chi}^{2}-2v^{2}) 
\end{pmatrix}.
\end{equation}

Since the VEV of fields equal their classical values, that implies $\langle \phi^{(\ast)}_{q} \rangle=0$ and $\langle \chi_{q} \rangle=0$ for quantum fields. Before calculating the quantum corrections by integrating-out quantum fields, we should establish appropriate power counting. In order to study Q-ball solutions, we assume the following mass hierarchy: $W_{\phi^{\ast}\phi} \ll W_{\chi \chi}$ (the opposite case was previously studied in \cite{Farhi_1998}). The simplest theory description is given by neglecting purely $\phi$ field terms \cite{PhysRevD.13.2739}. In this case, we set the mass matrix $W$'s power counting by assuming that couplings $h \sim z$ and $\varkappa \sim 1$, while $z\ll 1$. The mass matrix $W$ is a field-dependent construct, and by choosing relatively large fields values, we might break our mass hierarchy: $W_{\phi^{\ast}\phi} \ll W_{\chi \chi}$. Later, we will show that on the soliton configurations, this assumption is hold and we check that $\Hat{\chi} \lesssim 1$ and $|\Hat{\phi}| \ll \frac{\varkappa v}{h^{2}}$.

\begin{figure}[!tbp]
    \centering
    \includegraphics[width = 0.45\textwidth]{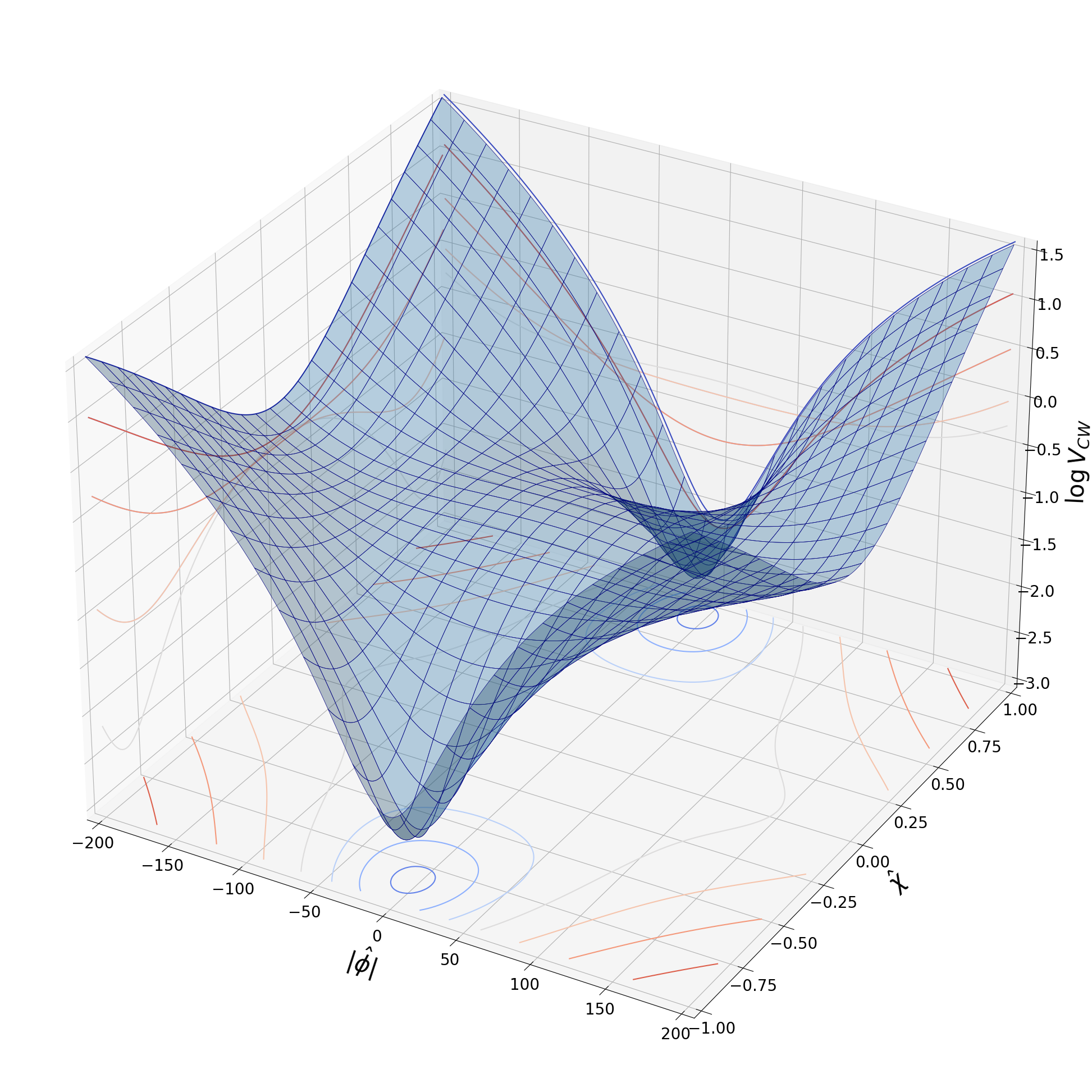}
    \caption{The 1-loop Coleman-Weinberg potential profile of the UV-completed FLS model. The parameters of the model are set to $h=0.01$ and $\varkappa = 1$.}
    \label{FLS_CW}
\end{figure}

In $(3+1)$ dimensions, nondimensionalization of the FLS model Lagrangian can be done by extracting the $v^{4}$ multilplier. Without loss of generality, we set $v=1$. The Coleman-Weinberg potential is represented as a perturbation series expansion in parameter $z$: $V_{CW}=\sum_{n=0}^{N} g_{n}(|\phi|,\chi) z^{n} $. In the leading order of perturbations, $\phi^{(\ast)}_{q} = 0$ and $\chi_{q}$ is integrated-out as

\begin{equation} \label{quant corr}
    V_{1-loop} = \frac{1}{64\pi^{2}}\left[ W_{\chi \chi}^{2}\left(\log{\frac{W_{\chi \chi}}{\mu_{H}^{2}}-\frac{3}{2}} \right) \right],
\end{equation}
where $\mu_{H}$ is an energy scale of order $\mu_{H}^{2}\sim W_{\chi \chi}$, and the 1-loop contribution is determined specifically at non-negative $W_{\chi \chi}$. Thus, we compute the Coleman-Weinberg potential at the renormalization scale associated with the heavy field $\chi$. By setting a field-dependent renormalization scale $\mu=\mu_{H}(|\Hat{\phi}|,\Hat{\chi})$ we are able to study quantum corrections even in the limit of small $W_{\chi \chi}$ \cite{Coleman:1973jx, Okane:2019npj}. The renormalized Coleman-Weinberg potential with the 1-loop contribution of field $\chi_{q}$ is

\begin{equation}\label{cw potential}
    V_{CW}(|\Hat{\phi}|,\Hat{\chi}) = h^{2}|\Hat{\phi}|^{2}\Hat{\chi}^{2} + \frac{\varkappa^2}{2}(\Hat{\chi}^{2}-v^{2})^{2} + V_{1-loop}.
\end{equation}

The quantum corrections tend to alter the behaviour of the FLS model potential, as can be seen in Fig.\ref{FLS_CW}. The previously observed flat direction along the $\chi=0$ line is now corrected to be slightly raised. A more illustrative interpretation of the influence of the quantum corrections on the FLS model is given after constructing a mechanical potential \re{mech_pot} from Eq.\re{cw potential}.

The consideration of the next-leading orders in the Coleman-Weinberg potential requires a consistent integrating-out of quantum fields $\phi^{(\ast)}_{q}$ and $\chi_{q}$. These computations are more complicated due to the presence of the second mass scale $\mu_{L}$ associated with the light field $\phi$. Thus, as explicitly discussed in \cite{Manohar:2020nzp}, a proper integrating-out of quantum fields $\phi^{(\ast)}_{q}$ requires renormalization group (RG) evolution to the scale $\mu_{L}$ and then computing a light field induced part of the Coleman-Weinberg potential. Additionally, if Lagrangian \re{UV FLS} is considered a high-energy theory at some renormalization scale $\mu_{0}$, then an EFT construction will require RG evolution to low-energies $\mu_{H}$ and $\mu_{L}$. Thereby, a study of the $\beta$-functions (App.\ref{beta}) of the UV-completed FLS model is vital.

\section{Quantum-corrected Q-ball}\label{eff.poten.}

\begin{figure}[!tbp]
  \centering
    \includegraphics[width = 0.45\textwidth]{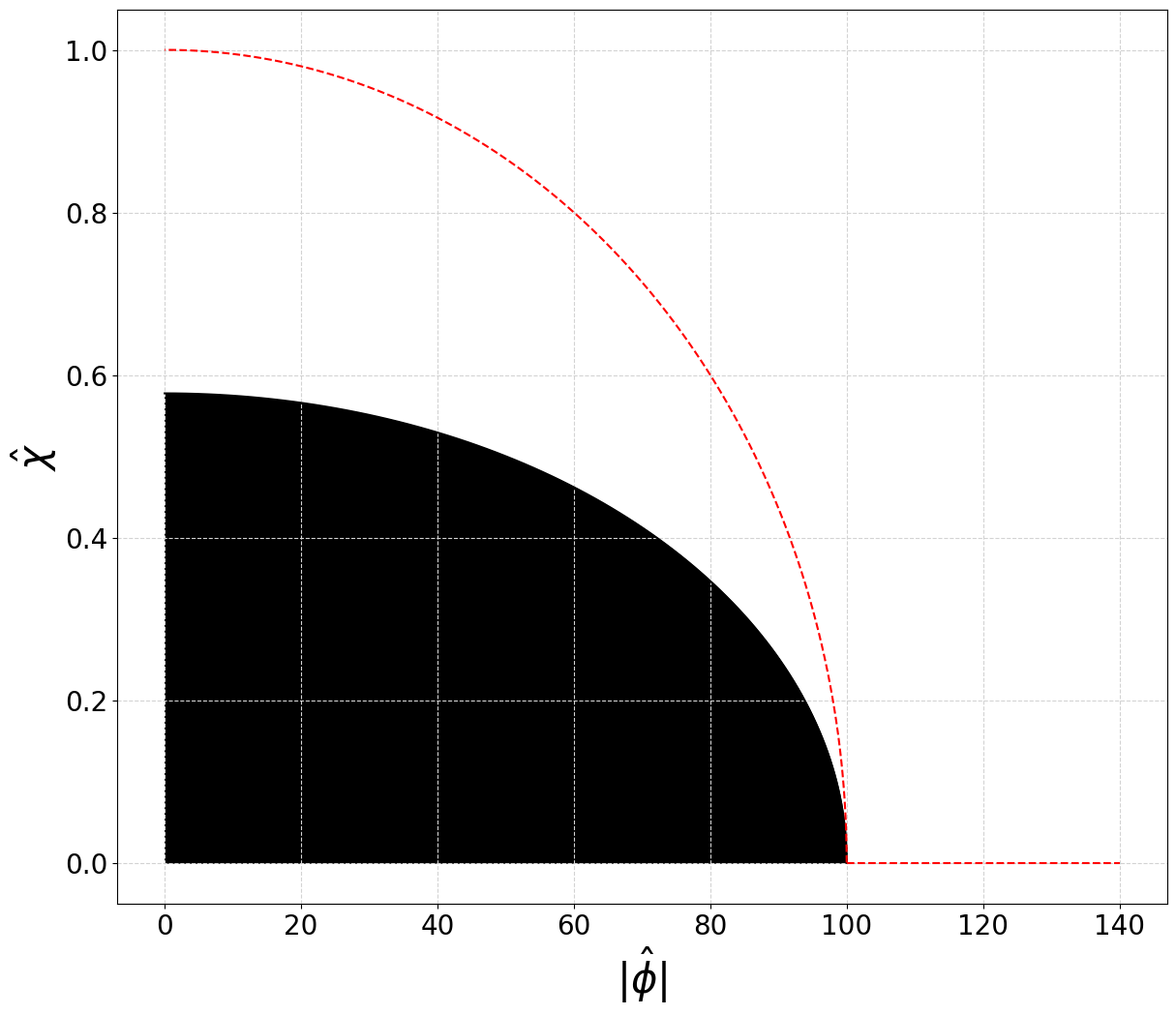}
    \caption{The black-filled region marks field values at which the $V_{1-loop}$ potential is not valuable. Classical-level integrating-out the field $\chi$ is shown in a dashed red line. The parameters are set to $h=0.01$ and $\varkappa = 1$.}
    \label{fig1}
\end{figure}

In the previous section, we introduced the quantum-corrected FLS model expressed in terms of the classical fields. Now we recall the analysis introduced in Sec.\ref{UV} to study the influence of the quantum corrections on the Q-ball solutions. Firstly, we integrate-out heavy field $\chi$ on the classical level by using equations of motion

\begin{equation}
     \frac{\partial V_{CW}(|\Hat{\phi}|,\Hat{\chi})}{\partial \Hat{\chi}} = 0.
\end{equation}

\begin{figure}[!tbp]
    \includegraphics[width = 0.45\textwidth]{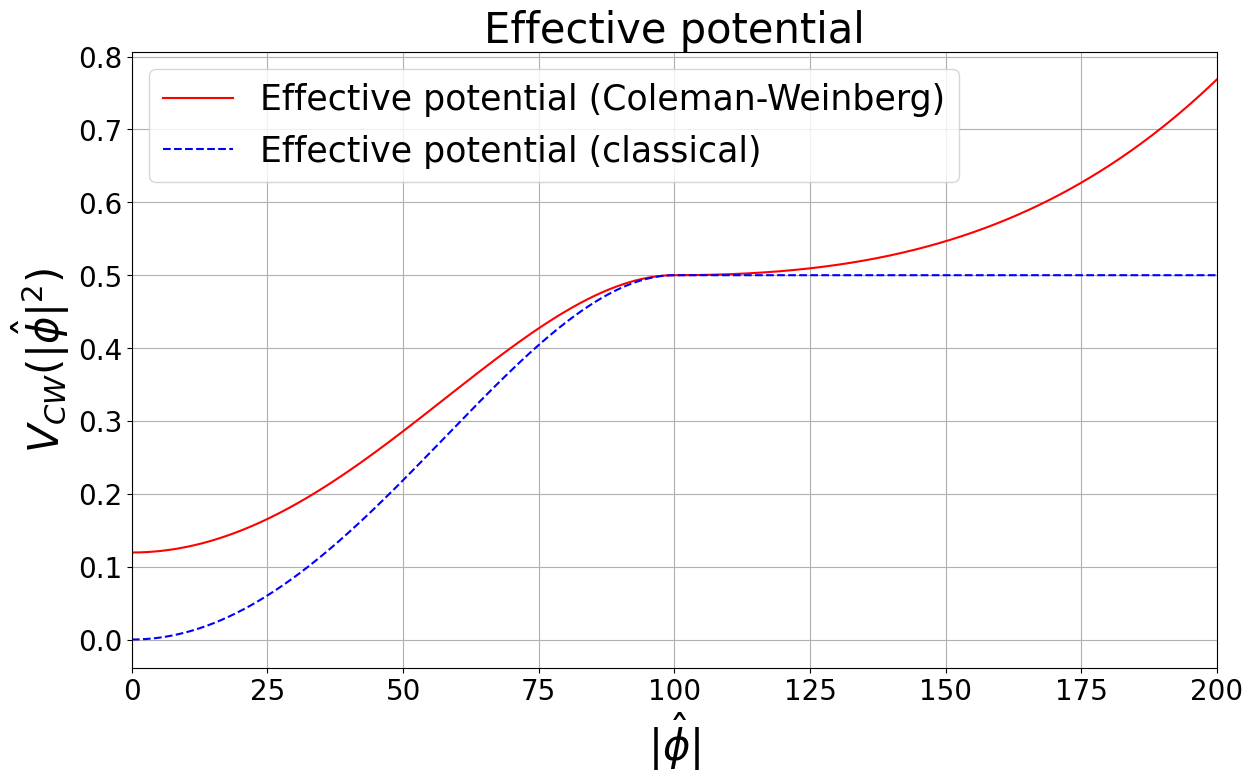}
    \caption{The effective potential $V_{CW}(|\phi|^{2})$ is shown. The parameters are set to $h=0.01$ and $\varkappa = 1$.}
    \label{fig2}
\end{figure}

As we decrease the value of the coupling $h$, the classical integrating-out procedure Eq.\re{int-out chi} becomes increasingly applicable. The field-space integrating-out trajectory is shown in Fig.\ref{fig1}. From Fig.\ref{fig2}, one can figure out that the effective potential is of the form 
\be
\begin{split}
&V_{eff}(|\Hat{\phi}|) = \left[m_{L}^{2}|\Hat{\phi}|^2 - \frac{\lambda_{L}}{2}|\Hat{\phi}|^{4} + \Hat{\Lambda}_{L} \right]\Theta\left(\frac{m_{L}}{\sqrt{\lambda_{L}}} -|\Hat{\phi}| \right) +\\
&+\left[-m_{R}^{2}|\Hat{\phi}|^2 + \frac{\lambda_{R}}{2}|\Hat{\phi}|^{4}+ \Hat{\Lambda}_{R} \right]\Theta\left(|\Hat{\phi}|-\frac{m_{L}}{\sqrt{\lambda_{L}}}\right),
\end{split}
\ee
where $\Hat{\Lambda}_{L,R}$ is a constant shift, and $\Theta(x)$ is a Heaviside step function. A simple condensate stability analysis by using Eq.(\ref{1-field instability}) shows that for large field $|\Hat{\phi}| > \frac{m_{L}}{\sqrt{\lambda_{L}}}$ values, condensate is stable. In this effective potential model, one can numerically look for Q-ball solutions. In this work, we have done numerical integration by Runge-Kutta of fourth order (lattice spacing $\epsilon \sim 10^{-1}$) with the shooting method of initial conditions. The quantum corrections of the classical FLS model (\ref{eq.1}) generate a $\lambda|\phi|^4$ term and result in the Q-ball's behaviour being similar to the common $|\phi|^{6}$ Q-ball model. 

\begin{figure}[!tbp]
    \centering
    \includegraphics[width = 0.45\textwidth]{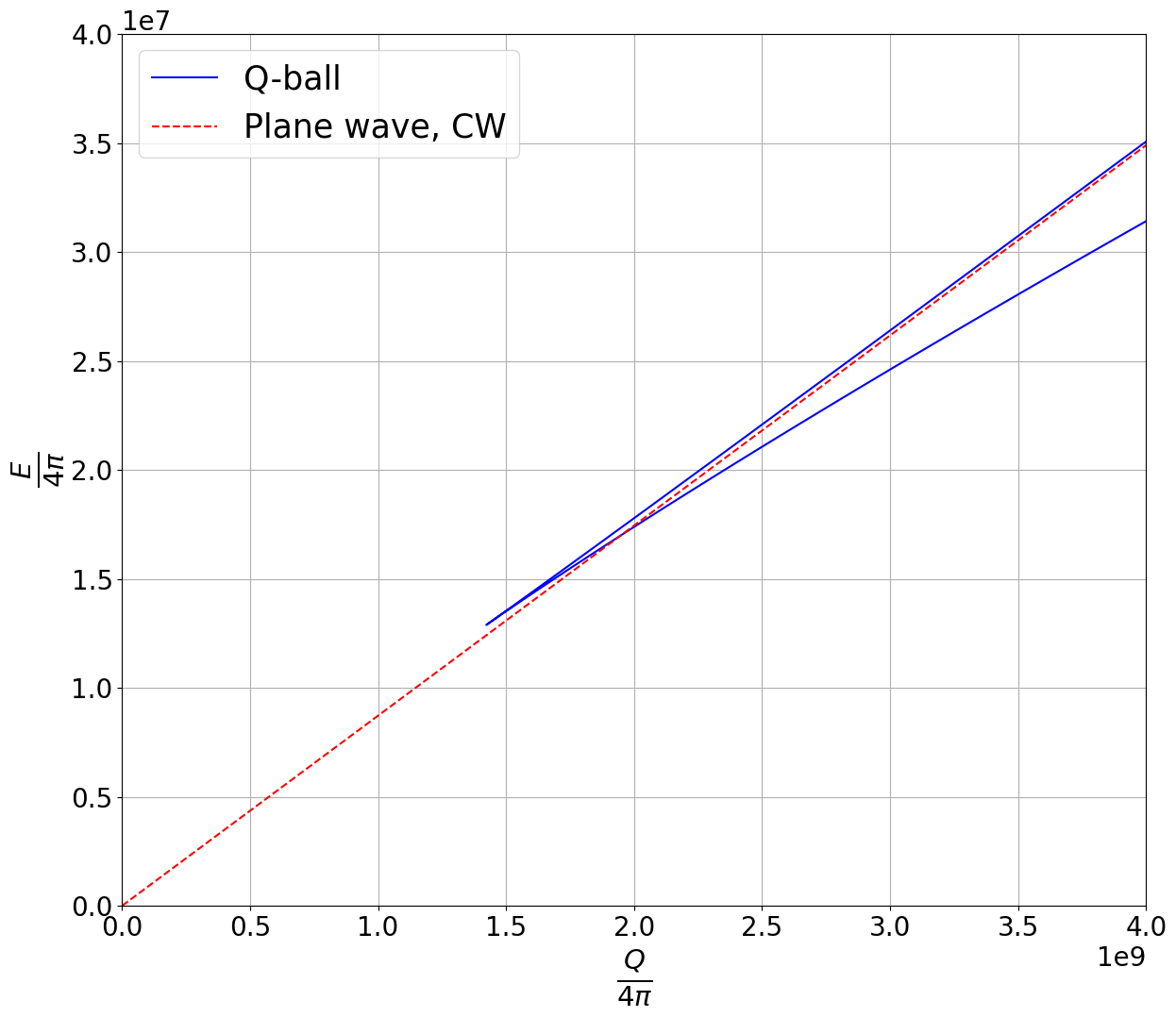}
    \caption{The Q-ball energy $\frac{E}{v}$ (regularized by subtracting $\Hat{\Lambda}_{L}$ from the effective potential to set $V_{eff}(0)=0$) is plotted as a function of $U(1)$ charge $Q$ for theory with effective potential. Parameters are set to $h=0.01$ and $\varkappa = 1$.}
    \label{fig3}
\end{figure}

After constructing an effective potential, we use redefined Eq.(\ref{mech_pot})
\be
U(|\Hat{\phi}|) = \omega^{2}|\Hat{\phi}|^{2} - V_{eff}(|\Hat{\phi}|) + \Hat{\Lambda}_{L},
\ee
so that $U(0)=0$. One can notice that in this model, Q-ball solutions possess the same type of behaviour that was shown in Sec.\ref{UV}. The lower bound on parameter $\omega$ is
\be
\omega_{-}^{2} = \sqrt{2\lambda_{R}(\Hat{\Lambda}_{R}-\Hat{\Lambda}_{L})}-m_{R}^{2}.
\ee
In the limit $\omega \rightarrow \omega_{-}$, the complex field $\phi$ solution profile resembles a thin-wall regime solution, and then finally settles at the condensate solution
\be
\Hat{\phi} = C e^{-i\omega_{-}t}, \text{ where } C = \sqrt{\frac{2(\Hat{\Lambda}_{R}-\Hat{\Lambda}_{L})}{\lambda_{R}}}.
\ee
Another peculiarity of the quantum-corrected Q-balls is an energy asymptotic at large $U(1)$ charge $E \propto Q$. More precisely, the energy functional in $(3+1)$ dimensions for spherically-symmetric solutions \cite{Nugaev:2019vru} can be written as 
\be
E = \omega Q + \frac{8\pi}{3}\int dr r^{2} \left[f^{'}(r)\right]^{2},
\label{eq.20}
\ee
which, along with a relation between $U(0)$ and $U(|\Hat{\phi}|_{in})$
\be
U(|\Hat{\phi}|_{in})-U(0) = 2\int dr \frac{\left[f^{'}(r)\right]^{2}}{r}
\label{eq.21}
\ee
shows that as $\omega \rightarrow \omega_{-}$, the r.h.s. of the Eq.\re{eq.21} vanishes. Thus, one can see from Eq.\re{eq.20} that the energy asymptotic emerges as
\be
E \propto \omega_{-}Q, \text{ as } Q\rightarrow \infty.
\ee
For the couplings $h=0.01$ and $\varkappa=1$, parameter $\omega_{-}$ takes value $\omega_{-}\approx 0.004 $.

\begin{figure}[!tbp]
    \centering
    \includegraphics[width = 0.45\textwidth]{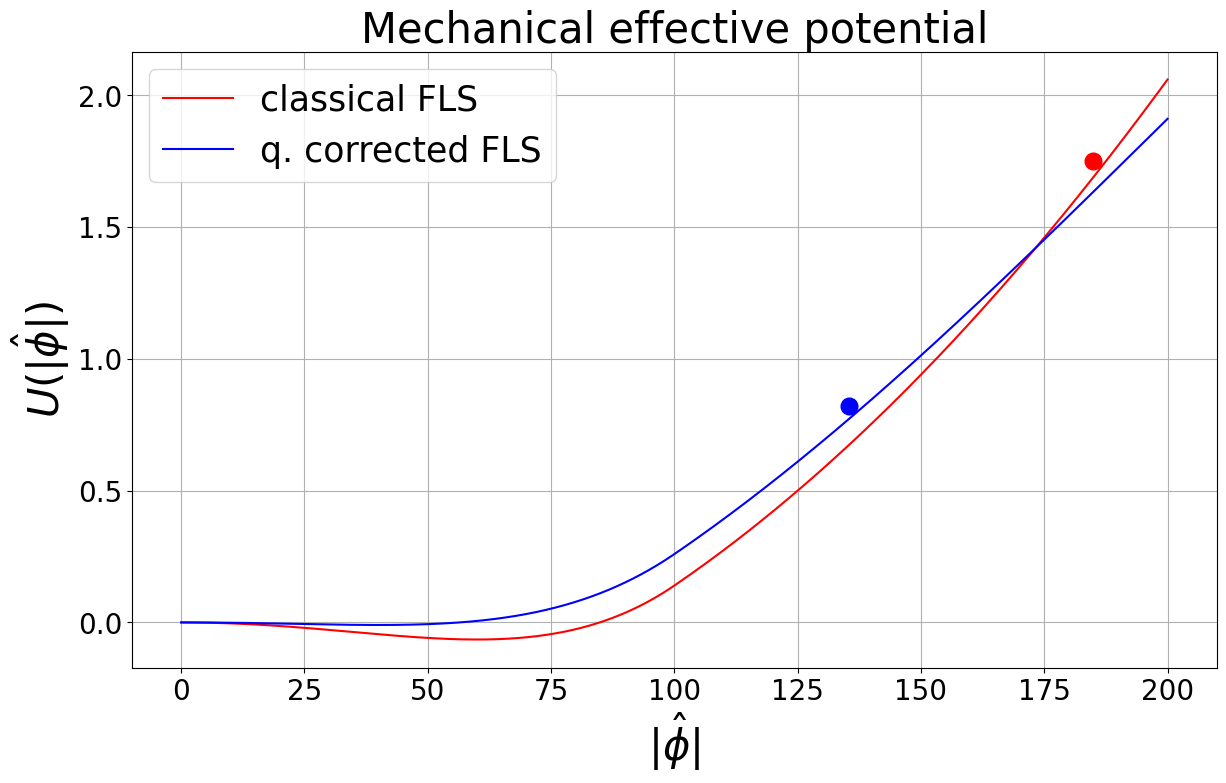}
    \caption{The comparison of the initial values $f(0)$ (presented as color-filled dots on the plot) of the Q-ball solutions for the classical/quantum corrected FLS models effective potential at $h = 0.01$, $\varkappa=1$ and $\omega = 0.8$.}
    \label{class quant}
\end{figure}

Lastly, it is worth providing a mechanical illustration (see Fig.\ref{FLS_mech_eff} and Fig.\ref{class quant}) of the influence of the quantum corrections on the Q-ball solutions. Despite bringing small corrections (see Eq.\re{quant corr}), the most change is due to the alteration of the flat direction of the classical FLS model, where any changes are very noticeable.

\section{Outlook}\label{outlook}

In this paper, we have considered the ultraviolet completion of the Friedberg-Lee-Sirlin model under the assumption of the $\chi\rightarrow -\chi$ symmetry (one can violate this symmetry to obtain a more general model). At the classical level, the behaviour of non-topological solitons in $(3+1)$ dimensions is quite different from the solutions of the common FLS model. The presence of a quartic self-interaction term for the complex scalar field results in the occurrence of a lower bound on the parameter $\omega$, defined as $\omega_{-}$ in Eq.\re{w_min_1}. In addition to this, as $\omega \rightarrow \omega_{-}$, a soliton is starting to resemble a flattened profile, while at $\omega=\omega_{-}$ the solution turns into a stable condensate. These characteristics of non-topological solitons in the UV-completed FLS model are similar to those observed for Q-balls in the $|\phi|^{6}$ model \cite{Mai:2012yc}. 

An advantage of having an ultraviolet-complete QFT model is the possibility to study the effect of quantum corrections on non-topological solitons. Under the assumption of a hierarchy of mass scales in the \re{UV FLS} model, we computed a 1-loop Coleman-Weinberg potential in leading order. On the renormalization scale associated with the mass of the heavy real field $\mu_{H}\sim m_{\chi}$, Coleman-Weinberg potential \re{cw potential} changes the potential of the FLS model \re{eq.1} as shown in Fig.\ref{FLS_CW}. By using the effective potential method we saw that quantum corrections generate a quartic self-interaction term $+\lambda|\phi|^{4}$. It is important to note that initially the parameter $\omega$ had no lower bound, but through the Coleman-Weinberg mechanism quantum corrections induce a new parameter $\omega_{-}$ with the dimension of mass. Quantum corrected non-topological solitons of the FLS model are of interest in various phenomenological models in cosmology, particle physics, etc. In contrary to the Q-balls \cite{Enqvist:1997si,Kusenko:1997si, Tsumagari:2009na}, Friedberg-Lee-Sirlin model does not require an extension to supersymmetry. However, as shown in this work, Coleman-Weinberg mechanism coupled with a classical effective potential method allows to consider a quantum-corrected Q-balls in various cosmological models \cite{ jiang2024gauged,Troitsky_2016,SPECTOR1987103}. 

Up to this point, we have discussed only the Coleman-Weinberg potential for the FLS-type model in flat space-time. Minimal coupling to gravity gives rise to the  boson stars, stationary solutions of the coupled system of the Einstein-matter field equations \cite{Kaup:1968zz, Ruffini:1969qy} which can be smoothly linked to the corresponding Q-balls in Minkowski spacetime. The dynamical evolution of the self-gravitating Einstein-FLS  boson stars is very different, there is always a minimal value of angular frequency $\omega_{-}$. Further, the parameters of the boson stars, when considered as functions of the frequency, follow a spiraling/oscillating pattern, with successive backbendings \cite{Friedberg:1986tq}.  One would expect that, as gravitational coupling remains relatively weak, the contribution of the Coleman-Weinberg potential may modify this picture. On the other hand, this mechanism allows dynamical generation of the Planck scale in  inflationary models with non-minimal coupling with gravity  \cite{Karam:2018mft}. Note that, the gravitational Coleman-Weinberg mechanism was studied recently in the model with two real scalar fields \cite{Alvarez-Luna:2022hka}, an interesting task is to investigate Coleman-Weinberg effect in the $R^2$-FLS-type model with non-minimal coupling to gravity.   
We hope to
address these questions in our future work.

\section*{Acknowledgements}
Authors are indebted to Yulia Galushkina for  the support on the early stages of the work. E.K. is grateful to Kirill Kondratenko and Oraz Anuaruly for helpful comments on the paper. Y.S. is grateful to Valery Kiselev for inspiring and valuable discussions.
Y.S. would like to thank the Hanse-Wissenschaftskolleg Delmenhorst for support and  hospitality. Analytical studies of this work were supported by the grant RSF 22-12-00215.

\appendix

\section{1-loop $\beta$-functions}\label{beta}

In a 1-loop approximation, the relation

\begin{equation}
    \beta_{i}\frac{\partial}{\partial \alpha_{i}}V(|\phi|^{2},\chi) = \frac{1}{2}\Tr{W^{2}},
\end{equation}
for couplings $\alpha_{i}$ defines corresponding $\beta$-functions \cite{Manohar:2020nzp}

\begin{subequations}\label{eqn}
    \begin{align*}
        & \beta_{m_{\phi}^{2}} = m_{\phi}^{2}\lambda - 4h^{2}v^{2}\varkappa^{2}, \tag{\ref{eqn}} \\
        & \beta_{\lambda} = 2\lambda^{2} + 8h^{4}, \\
        & \beta_{h^2} = h^{2}\lambda + 12h^{2}\varkappa^{2} + 4h^{4}, \\
        & \beta_{m_{\chi}^{2}} = -2h^{2}m_{\phi}^{2}+24v^{2}\varkappa^{4}, \\
        & \beta_{\varkappa^{2}} = h^{4} + 36\varkappa^{4}, \\
        & \beta_{\Lambda} = \frac{1}{2}m_{\phi}^{4} + 2\varkappa^{4}v^{4},
    \end{align*}
\end{subequations}
where the potential in \ref{UV FLS} was rewritten as

\begin{equation}
    V(|\phi|^{2},\chi) = m_{\phi}^{2}|\phi|^{2} + \frac{\lambda}{4}|\phi|^{4} + h^{2}|\phi|^{2}\chi^{2} - \frac{m_{\chi}^2}{2}\chi^{2}+ \frac{\varkappa^2}{2}\chi^{4} + \Lambda.
\end{equation}
As we see from the equations above, the FLS model without counterterms is not properly renormalizable.

\bibliographystyle{elsarticle-num} 
\bibliography{example}






\end{document}